\documentclass[preprintnumbers,amssymb,12pt]{revtex4}
\begin{document}
\newcommand{\im}{{\rm Im}}
\newcommand{\sech}{{\rm sech}}
\newcommand{\kp}{{K\"ahler potential }}
\newcommand{\csch}{{\rm csch}}
\newcommand{\beq}{\begin{eqnarray}}
\newcommand{\eeq}{\end{eqnarray}}
\newcommand{\nn}{\nonumber}
\newcommand{\hc}{{\rm h.c.}}
\newcommand{\La}{\mathcal{L}}
\def\ltap{\ \raise.3ex\hbox{$<$\kern-.75em\lower1ex\hbox{$\sim$}}\ }
\def\gtap{\ \raise.3ex\hbox{$>$\kern-.75em\lower1ex\hbox{$\sim$}}\ }
\def\CO{{\cal O}}
\def\CL{{\cal L}}
\def\CM{{\cal M}}
\def\tr{{\rm\ Tr}}\
\def\CO{{\cal O}}
\def\CL{{\cal L}}
\def\CM{{\cal M}}
\def\tr{{\rm\ Tr}}
\newcommand{\bel}[1]{\be\label{#1}}
\def\al{\alpha}
\def\bt{\beta}
\def\eps{\epsilon}
\def\mn{{\mu\nu}}
\newcommand{\rep}[1]{{\bf #1}}
\def\be{\begin{equation}}
\def\ee{\end{equation}}
\def\bea{\begin{eqnarray}}
\def\eea{\end{eqnarray}}
\newcommand{\eref}[1]{(\ref{#1})}
\newcommand{\Eref}[1]{Eq.~(\ref{#1})}
\newcommand{\gsim}{ \mathop{}_{\textstyle \sim}^{\textstyle >} }
\newcommand{\lsim}{ \mathop{}_{\textstyle \sim}^{\textstyle <} }
\newcommand{\vev}[1]{ \left\langle {#1} \right\rangle }
\newcommand{\bra}[1]{ \langle {#1} | }
\newcommand{\ket}[1]{ | {#1} \rangle }
\newcommand{\ev}{ {\rm eV} }
\newcommand{\kev}{ {\rm keV} }
\newcommand{\Mev}{ {\rm MeV} }
\newcommand{\gev}{ {\rm GeV} }
\newcommand{\tev}{ {\rm TeV} }
\newcommand{\mev}{ {\rm meV} }
\newcommand{\ma}{m^2_{\rm atm}}
\newcommand{\ml}{m^2_{\rm LSND}}
\newcommand{\tl}{\theta_{\rm LSND}}
\newcommand{\ms}{m_\odot^2}
\newcommand{\cta}{c_{\rm a}}
\newcommand{\cts}{c_\odot}
\newcommand{\sta}{s_{\rm a}}
\newcommand{\sts}{s_\odot}
\newcommand{\ctm}{c_{\rm m}}
\newcommand{\stm}{s_{\rm m}}
\newcommand{\mpl}{M_{Pl}}

\title{Little Inflatons and Gauge Inflation}

\author{David E. Kaplan$^{a,c}$\footnote{dkaplan@pha.jhu.edu}}
\author{Neal Weiner$^{b,c}$\footnote{nealw@phys.washington.edu}}
\affiliation{$^a$ Department of Physics and Astronomy, Johns Hopkins
University, Baltimore, MD  21218, USA}
\affiliation{$^b$ Department of Physics, University of Washington,
Seattle, WA 98195-1560, USA}
\affiliation{$^c$ Department of Physics, Boston University, Boston, MA
02215}
\date{\today}
\preprint{UW/PT 03-03}
\preprint{BUHEP-03-02}

\begin{abstract}
Cosmological inflation gives a natural answer for a variety of
cosmological questions, including the horizon problem, the flatness
problem, and others. However, inflation yields new questions relating to
the flatness of the inflaton potential. Recent studies of ``little''
fields, a special class of pseudo-Goldstone bosons, have shown it is
possible to protect the mass of a field while still yielding order one
interactions with other fields. In this paper, we will show that ``little
inflatons'' are natural candidates for the slow roll field of hybrid
inflation models. We consider both supersymmetric and non-supersymmetric
models, and give a simple examples based on approximate Abelian symmetries
which solve the inflaton flatness problem of supergravity.  
We also present hybrid models in which components of gauge 
fields in higher dimensions play the role of the
inflaton.  Protected by higher-dimensional gauge symmetry, they, too, naturally
have large couplings while suppressed mass terms. We summarize the implications of the new WMAP data on such models.
\end{abstract}
\maketitle

\newpage
\section{Introduction}
Inflation \cite{Guth:1981zm} is an important feature of most modern
cosmologies. It not only provides solutions for the horizon problem and
the
flatness problem, it also provides a natural explanation for the spectrum
of density
perturbations observed by COBE \cite{Bennett:1996ce} and the more recent
experiments such as
MAXIMA \cite{Lee:2001yp}, BOOMERANG \cite{deBernardis:2001xk} and DASI
\cite{Kovac:2002fg}. Even more recently, the WMAP \cite{Bennett:2003bz,Spergel:2003cb}
experiment has added a great deal of new precision data, 
giving great new information into the physics of inflation.

In building models of inflation, one generally requires three basic
ingredients: a
large background energy density to drive inflation, a scalar field $\phi$
which will roll from some initial value to some other value where
inflation
ends, and the requirement that the energy associated with the kinetic
energy is
subdominant to the background energy density (the slow-roll conditions).
With
recent measurements, one can also constrain the form of the inflaton
potential
by studying density perturbations.

The simplest implementation of chaotic inflation
\cite{Linde:1983gd} is just a scalar field with a polynomial
potential. However, when the slow roll conditions (below) are
applied, these models typically require super-Planckian vevs to drive
inflation.  In other words, because the potential is flat, one
must start at very large field values to produce enough vacuum
energy.

One interesting solution, hybrid inflation \cite{Linde:1994cn},
represented significant progress in inflation model-building.   If
one field is responsible for slowly rolling while a different
field is kept from relaxing to its minimum, then it is possible to
divorce the size of the vacuum energy from the slow roll potential
in a relatively natural way. In its original form, there are two
fields, $\phi$, the slow-roll field and $\sigma$, the
``waterfall'' field, and a potential
\be
V(\phi,\sigma) = \lambda (M^2- \sigma^2)^2 + \frac{m^2}{2}
\phi^2 + \frac{g^2}{2} \phi^2 \sigma^2.
\label{eq:hybrid}
\ee
When the slow roll field has a large value, $\sigma$ has a large positive
mass squared and is quickly driven to zero. From this point the universe
inflates with an energy density $\sim \lambda M^4$ and $\phi$ 
slowly rolls until it reaches $\phi_c\equiv \sqrt{4\lambda M^2 / g^2}$ 
where the mass squared of $\sigma$
changes sign.  At this point inflation ends.

The slow roll requirements are $\epsilon \ll 1$ and $|\eta| \ll 1$ where
\bea \epsilon &\equiv&
\frac{\mpl^2}{2}\left( \frac{V'}{V} \right)^2 , \\ \eta
&\equiv& \frac{\mpl^2 V''}{V}. \eea 
These imply that
$m^2 \ll \lambda M^4/\mpl^2$ (where $\mpl \simeq 2\times 10^{18}$
GeV, the reduced Planck mass).  This introduces a significant
model building problem: the (squared) mass of a scalar
field receives quadratically divergent 
quantum corrections generically making it too large and making it sensitive
to physics at the ultraviolet cutoff of the theory.

A technically natural mass for $\phi$, taking the cutoff to be
$\mpl$, is $m^2\sim (g/4\pi)^2 \mpl^2$.  With an $m^2$ about a
tenth this size (requiring $\sim 10\%$ fine-tuning) it is possible
to generate enough e-foldings of inflation
\begin{eqnarray}
N(\phi_i ) & = & {1\over \mpl^2} \int_{\phi_c}^{\phi_i} {V\over
V'} d\phi\\ & = & \int_{\phi_c}^{\phi_i} \frac{1}{\eta} {d\phi
\over \phi} \gsim 60
\end{eqnarray}
and density perturbations consistent with the COBE data:
\be
V^{1/4} = 6 \epsilon^{1/4} \times 10^{16} {\rm GeV}.
\ee
However, the couplings in this model are $\lambda \sim 10^{-11}$
and $g^2 \sim 10^{-10}$.  In addition, the vevs are
$M,\phi_c\sim\mpl$ thus requiring all higher-dimensional operators
to be suppressed by similar factors.  Such extreme values can be 
avoided at the cost of dramatically increasing the fine-tuning required.

There are only two known symmetries which can protect a scalar field from
acquiring a mass: supersymmetry and the shift symmetry of a Goldstone boson.
Supersymmetry would seem a natural choice because of the ubiquitous flat
directions in supersymmetric theories, but inflation is driven by a vacuum
energy density which breaks supersymmetry and generically gives
Hubble-scale masses to all scalar fields in the theory \cite{Copeland:1994vg}.

The shift symmetry of a Nambu-Goldstone boson (NGB) will, indeed, protect
it from acquiring a mass, even during inflation. However, the inflaton 
cannot be an exact NGB
as this same shift symmetry would prohibit the coupling required in the 
hybrid model.  Pseudo Nambu-Goldstone bosons (pNGB) have couplings 
which violate the shift symmetry by small amounts and thus these 
fields remain light.  However, simply calling
$\phi$ a pNGB is not enough to correct the above picture - the parameter
$m^2$ still gets a one-loop quadratic divergent contribution from $\phi$'s 
coupling to $\sigma$ which puts strain on the model.  
It is this contribution which must be suppressed.

The first use of pNGBs for the inflaton was so-called  
``Natural Inflation'' \cite{Freese:1990rb,Adams:1993bn}
(see also \cite{Frieman:1995pm}) in which a pNGB sat at a 
particular point in its potential from whence it could roll.
They have also been used, for example, as the waterfall field in 
supersymmetric hybrid inflation models \cite{Randall:1996dj}.  
However, these models did not suitably take into account possible 
gravitational corrections, which can be significant. Indeed, 
in \cite{Randall:1996dj}, the slow roll field would still naturally 
pick up a large mass from gravitational interactions. 

Recently, a new class of pNGBs have been discovered
\cite{Arkani-Hamed:2001nc} in which loop contributions to masses
are suppressed.  The suppression is due to the ``non-local'' or ``collective''
behavior of the explicit symmetry breaking:  the breaking comes from
a combination of couplings in the Lagrangian.
The result is a vanishing tree-level mass and
the cancelation of quadratic divergences at one or more loops.  
The first application of this mechanism was to the question of electroweak
symmetry breaking \cite{Arkani-Hamed:2001nc}, where the Higgs
quartic arose from multiple symmetry breaking terms and the loop
contributions to its mass were suppressed.
A ``Little Inflaton'' theory is therefore one in which the slow-roll field
$\phi$ is a little pNGB.  It can have non-trivial interactions with
the waterfall field (which itself may or may not be a pNGB)
without generating large corrections to its mass.  

We will show that the use of the above mechanism will maintain
a flat potential for the slow-roll field against all dangerous corrections,
including gravitational.  In the next section
we review the mechanism of collective symmetry breaking (CSB) and then
present a little inflaton model.

In Section \ref{sec:susy} we review the difficulty in realizing
inflation models in supersymmetric theories coupled to
supergravity (the ``$\eta$ problem'').  We then present a simple
supersymmetric model using the same mechanism as in the previous
case.  The dimensionless parameters are all order one and
the model only requires one additional scale besides the Planck scale.
We include a sequestered model in an extra dimension which gives a
natural microscopic explanation for the form of the potential.

Another naturally light scalar with a protective shift symmetry is
the fifth component of a gauge field in an extra dimension.
In Section \ref{sec:gauge} we present models of ``gauge
inflation'' in which the slow roll field is the fifth component of
a $U(1)$ gauge field, again achieving natural inflation with order
one parameters.  One can see from dimensional deconstruction 
\cite{Arkani-Hamed:2001ca} that these extra components of the gauge
field are akin to pNGBs.  We discuss the deconstruction of a particular 
six-dimensional model of this type.

In Section \ref{sec:WMAP}, we discuss the significance of the recent WMAP results on the given models. Finally, in Section \ref{sec:conc} we conclude.

\section{Goldstone Bosons, Collective Breaking and a New Model of Inflation}
\label{sec:littlest} 
Before discussing explicit models of
inflation, we describe a new mechanism which leads to a naturally
flat scalar potential.  Until now, this mechanism has only been
applied to electroweak symmetry breaking to produce a light Higgs
boson mass compared to the cutoff of the theory.  Here we apply it
to the inflaton potential.

A NGB is a scalar field that results from a spontaneous broken
global symmetry.  For example, we can consider an $SO(2)$ symmetry broken
by a linear sigma model field
\begin{eqnarray}
\Phi= e^{i \tau_2 \theta(x^\mu) / f}
    \left( \begin{array}{c} 0 \\ f + \rho(x^\mu) \end{array} \right)
\end{eqnarray}
(where the normalization is chosen for notation ease later). It will
frequently be
convenient to set $\rho=0$, that is to consider the non-linear sigma model
in which
we do not discuss the dynamics of symmetry breaking. If $\Phi$ appears
only in terms
which respect the $SO(2)$ symmetry, the field $\theta$ will have only 
derivative interactions.

A pseudo- or approximate NGB is one in which the potential explicitly
breaks the global symmetry with small couplings.  An example of
pseudo-NGBs
in nature are the pions of QCD.  Quark masses and electric charge break
the
chiral symmetry explicitly and give the pions masses at tree-level and one
loop respectively.

The inflaton is a natural candidate for a NGB as we have discussed,
because of the
necessary flatness of the potential. However,
calling $\phi$ in the hybrid inflation model above a pNGB doesn't change
the theory - specifically, $\phi$ still receives a large mass term
via quantum effects.

The problem with the hybrid model considered as theory of a pNGB inflaton
is that a single coupling in the potential broke the symmetry
protecting $\phi$ from a mass term. A quadratically divergent
renormalization was thus allowed to appear at one-loop, which is generically too large.

However, if the global symmetry is explicitly broken by a combination of
couplings, then {\em loop contributions to pNGB masses must involve
all of the couplings, and the one-loop contribution cannot be
quadratically divergent}.  This is the essence of collective
or non-local symmetry breaking \cite{Arkani-Hamed:2001nc}.  The
effect is to suppress loop contributions to the $\phi$
mass in the model of the previous section, naturally making
the slow-roll parameter $\eta$ small.

\subsection{The littlest inflaton}
We begin by considering the simplest model of the type described above.
It involves replacing $\phi$ with a pseudo-NGB $\theta$ which comes from
the breaking of a global $SO(2)$ symmetry.  We integrate out the
``radial'' degree of freedom $\rho$ and push the cutoff of this non-linear sigma
model to the point where $\phi$ becomes strongly coupled, namely
$\Lambda\sim 4\pi f$.  Thus the inflaton is parameterized as
\begin{eqnarray}
\Phi = \left(\begin{array}{cc} \cos{(\theta/f)}&\sin{(\theta/f)}\\
-\sin{(\theta /f)}
    & \cos{(\theta /f)}\end{array} \right) \left(\begin{array}{c} -1\\
1\end{array}\right)
    \times {f\over\sqrt{2}}
\end{eqnarray}
We choose to parameterize the $SO(2)$ breaking in the direction above for
convenience later.
We take its tree-level potential to be:
\begin{equation}
V= \lambda \left( \sigma^T \sigma - v^2 \right)^2
+ {g_1\over 4} \,(\sigma^T {\Phi})^2
+ {g_2 \over 4} \,(\sigma^T \tau_1 \Phi)^2
\end{equation}
where $\sigma^T = (\sigma_1 \, \sigma_2)$ and $\tau_1$ is the first Pauli
matrix.

When $g_1 = g_2 = 0$, $\theta$ transforms non-linearly under an exact
$SO(2)$
symmetry while $\sigma$ transforms linearly under a {\it different}
$SO(2)$ symmetry.
Making, for example, $g_1$ non-zero explicitly breaks the two symmetries
to the diagonal
$SO(2)$ which leaves $(\sigma_1\, \sigma_2) \cdot \Phi$ invariant.  The
remaining $SO(2)$
is still spontaneously broken and thus $\theta$ still represents an exact
NGB.  To see this,
one can do a field redefinition which makes ${\tilde\sigma}^T = \sigma^T
e^{-i \tau_2 \theta / f}$,
which turns the $g_1$ operator into a mass term for the linear combination
$({\tilde\sigma}_1 - {\tilde\sigma}_2)$.  The transformation changes the
derivative
operators in the theory ({\it e.g.}, the kinetic terms for $\sigma$
produce higher-dimensional
operators), but the potential is independent of $\theta$.

If instead we make $g_2$ non-zero, we still leave an $SO(2)$ symmetry
preserved
under which $(\sigma_2\, \sigma_1) \cdot\Phi$ is a singlet.  We could
again go to
``unitary gauge'' and transform $\theta$ away.  However, with both terms
present,
we can only eliminate {\em one} term by a field transformation, leaving a
non-trivial
potential for $\theta$. However, since this potential and consequently the
renormalization
of the mass will require both $g_1, g_2 \neq 0$, the slow-roll conditions
are easily accommodated.

For the rest of this discussion we take $g_1=g_2 \equiv g$ for simplicity.
This can
be enforced by imposing a ${\cal Z}_2$ symmetry which takes
$\sigma_1\leftrightarrow \sigma_2$.
This simplification is not necessary for a successful model and the
dynamics are not dramatically
different in the more general case.  In addition, we have imposed a
$\sigma\rightarrow -\sigma$
symmetry on the Lagrangian to avoid terms linear in $\sigma$.

Now we compute the one-loop corrections to the mass of $\theta$.
From expanding out the $\Phi$s
in the potential,
\be
V = {g f^2\over 4} (\sigma_1^2 + \sigma_2^2 -
2\sigma_1\sigma_2\cos{(2\theta /f)} )
\ee
it is clear that there is no one-loop quadratic divergent contribution to
a $\theta$ mass.
This is because $\theta$ only couples to the combination
$\sigma_1\sigma_2$ making it impossible
to close a loop with only one vertex.  There is a logarithmic divergence
at one loop proportional to
$g_1 g_2 = g^2$
\begin{eqnarray}
V_{1-loop} &=&  \frac{g^2}{128\pi^2} \log{\left({\Lambda^2 \over
m_{\theta}^2}\right)} (\Phi^T \tau_1 \Phi )^2 + ...\nonumber\\ \nonumber\\
&=& \frac{g^2 f^4}{128\pi^2} \log{\left({\Lambda^2 \over
m_{\theta}^2}\right)} \cos^2{(2\theta/f)} + ...
\end{eqnarray}
Because this theory is a non-linear sigma model, it becomes strongly
coupled at a scale $\Lambda\simeq 4\pi f$.  We take this to be the cutoff of the
theory but for now leave its value unspecified.  It could be at $\mpl$ or far below.

Let us now compute the number of e-foldings.  For simplicity, we take the
VEV
of the waterfall field $\sigma$ to be $v = f \simeq \Lambda/ 4\pi$.  This
value is not
only technically natural, but the theory remains perturbative (and
therefore under control)
in this regime.  To find the critical value of $\theta$, we diagonalize
the $\sigma$ mass matrix:
\begin{eqnarray}
& &\sigma^T{g\over 4}  \left( \begin{array}{cc} 1 & \cos{(2\theta /f)} \\
\cos{(2\theta /f)} & 1 \end{array} \right) \sigma \nonumber\\ \nonumber\\
&\rightarrow&  {\tilde\sigma}^T{g\over 4} \left( \begin{array}{cc} 1 -
\cos{(2\theta /f)}
& 0 \\ 0 & 1 + \cos{(2\theta /f)}  \end{array} \right) {\tilde\sigma}
\end{eqnarray}
This gives us a value for $\theta_c$ satisfying
\begin{equation}
{4 \lambda\over g} = 1 - \cos{(2\theta_c /f)} =2
\sin^2{\left({\theta_c\over f} \right)}.
\label{pseudo-c}
\end{equation}
The $\sigma$ field is stabilized when $|\theta| > |\theta_c|$.
Note the sign of the mass term for $\theta$ needs to be opposite
that which appears in the loop calculation.  However, the sign is
not a prediction of the theory as there is a counterterm canceling
this logarithmic divergence at one loop (along with the two-loop
quadratic divergence).  We assume the sign is what we need for the
proper dynamics while the magnitude is approximately that
calculated above, {\it i.e.},
\begin{equation}
V_{\theta - mass} \simeq - \frac{g^2 f^4}{128\pi^2} \log{\left({\Lambda^2
\over m_{\theta}^2}\right)} \cos^2{(2\theta/f)} \equiv -\frac{\bar
g^2}{128 \pi^2} f^4 \cos^2{(2\theta/f)},
\end{equation}
where $\bar g^2 = g^2 \log(\Lambda^2 / m_\theta^2)$.
The number of e-foldings can be computed
\begin{eqnarray}\label{eq:e-folds}
N(\theta_i ) &=& {1\over \mpl^2}
    \int_{\theta_c}^{\theta_i} \frac{- 128 \pi^2 \lambda f^4 }{2 \bar g^2
f^3 \sin{(4\theta/f)}} d\theta \nonumber\\    &=& -\frac{\lambda}{\bar g^2} \log \left( \frac{ \tan(2
\theta_c/f)}{\tan (2 \theta_i/f)} \right)\\
    \nonumber
    &\sim& \frac{ \lambda  }{2 \bar g^2} \log(g/\lambda)
\end{eqnarray}
where we have assumed $\theta_c \ll f$ and we've taken
$\Lambda=\mpl = 4\pi f$.  Thus we see from eq. (\ref{pseudo-c}) that
$\lambda < g$ and from eq. (\ref{eq:e-folds}) we require $\lambda \gg g^2$.

COBE requirements on the spectrum imply
\be
\frac{ \lambda^{3/2} }{\pi \bar g^2  \sin (4 \theta/f)} = 5.2 \times
10^{-4}
\ee
or, roughly,
\be
\bar g^2 \approx \ \lambda^{3/2} \times 10^3.
\ee
Altogether, this implies $\bar g <  10^{-3}$ and $\lambda <
10^{-6}$. Numerically, then, we have $m_\sigma^2 \approx 4 \lambda
f^2 < (4 \times 10^{14} \gev)^2$ and $m_\theta^2 \approx \bar g^2 f^2/16 \pi^2 <
(2 \times 10^{13} \gev)^2$.

Thus, as an illustration of the mechanism,
we have shown an example of a hybrid model with only one additional
coupling ($g\rightarrow g_1,g_2$) which is technically natural, does not require
cutoff-size vevs, and yet has a much smaller inflaton mass than would be otherwise
expected.


\section{A Supersymmetric Little Inflaton}
\label{sec:susy}

Supersymmetry would seem a natural resolution to the flatness problem of the
inflaton potential. After all, scalar fields in supersymmetric theories
are protected by non-renormalization theorems, and the abundant flat directions in supersymmetric theories seem natural candidates for the inflaton field.

However, since inflation necessarily involves gravity, it is necessary to
embed the theory in supergravity. Because inflation occurs as a result of energy
density, supersymmetry is broken during inflation and fields generically
pick up masses of order the Hubble parameter, in direct conflict with slow-roll
requirements \cite{Copeland:1994vg}.

Let us see explicitly how this happens: the potential for a scalar field during
inflation is
\bea
V&=&\exp\left( \frac{K}{\mpl^2} \right)\times \\
\nonumber &&\left[ \sum_{\alpha, \beta} \left( \frac{\partial^2
K}{\partial \bar
    \phi_\alpha \partial \phi_\beta} \right)^{-1} \left( \frac{\partial
    W}{\partial \phi_\alpha}+\frac{W}{\mpl^2} \frac{\partial K}{\partial
    \phi_\alpha} \right)
\left( \frac{\partial W}{\partial \bar \phi_\beta}+\frac{W}{\mpl^2}
\frac{\partial K}{\partial \bar \phi_\beta} \right) -3
\frac{|W|^2}{\mpl^2}
\right],
\eea
where $W$ is the superpotential and $K$ is the K\"ahler potential.
Let us assume that the inflaton field has canonical \kp, and that there is a
source of vacuum energy, and thus supersymmetry breaking, in the 
$F$ term of some field, such that
$F^2 = \mu^4$. Then the inflaton field automatically picks up a mass through
supergravity couplings,
\be
V = \exp\left( \frac{K}{\mpl^2}  \right) (\mu^4+...).
\ee
For a canonical \kp $K(\phi^*, \phi) = \phi^* \phi$, this yields a
slow-roll parameter $\eta=1$ at tree level, absent fine tunings of other
terms in the potential.

There are at least two other approaches two address this: first of all, if
the slow-roll field enters only {\em linearly} in the superpotential,
there is a cancellation between this term and other terms
\cite{Linde:1997sj}. However, this requires no additional Planck-scale
vevs in the theory, such as dilaton vevs or moduli vevs -- a highly restrictive
assumption about the ultraviolet theory. Another
alternative is D-term inflation \cite{Binetruy:1996xj}, where the vacuum
energy is driven by a Fayet-Iliopoulos term, although the scales in such
models are often too high and it is difficult to avoid generating large $F$ terms.

\subsection{The Model}

As we have already described, a little pNGB is one which is a
pseudo-goldstone boson, whose mass is protected by a combination of
symmetries. In non-supersymmetric theories, this is typically realized
with
multiple approximate internal symmetries. In supersymmetric theories,
masses are
already protected by supersymmetry. For a pNGB to acquire a mass, only
terms
which combine both violations of supersymmetry as well as violations of
the
approximate internal symmetry can contribute.

A number of authors have previously considered models of pNGBs in supersymmetric theories. In \cite{Copeland:1994vg}, it was noted that pNGBs would not suffer from the
supergravity $\eta$
problem. Models with discrete non-Abelian symmetries were constructed by
\cite{Cohn:2000hc}. In \cite{Murayama:1994xu} such form of the \kp was invoked, while \cite{Adams:1997yd} noted that moduli fields can naturally yield appropriate K\"ahler potentials.  The authors of \cite{Kawasaki:2000yn,Yamaguchi:2001pw} considered
many models of pNGBs which have nearly exact symmetries except for small breaking
parameters. In \cite{Randall:1996dj}, pNGBs were considered as candidates for waterfall fields.

Goldstone bosons are protected by a shift symmetry, rather than
supersymmetry. Hence, even in the presence of a cosmological constant,
Goldstone
bosons in supergravity are massless. Let us understand this more
carefully. Suppose we have a field which spontaneously breaks a global $U(1)$ symmetry,
\be
W_{gb}=S (\Phi \bar \Phi - v^2).
\ee
$F$-flatness is achieved when
\bea
\phi= v e^{\theta/\sqrt{2}v}, \\
\bar \phi = v e^{-\theta/\sqrt{2}v}, \\
\eea
where $\theta$ is a complex scalar field. We can promote $\theta$ to a
superfield
$\Theta$, in which case the kinetic terms are
\be
K(\Theta,\Theta^\dagger)=\Phi^\dagger \Phi + \bar \Phi^\dagger \bar \Phi
=2 v^2
\cosh (\frac{\Theta + \Theta^\dagger}{\sqrt{2}v}).
\ee
Let us then as before assume the presence of a field with a non-vanishing
$F$-term. Then
the the potential for $\theta$ in the presence of supergravity is
\be
V=\exp \left[ \frac{2v^2}{\mpl^2}\cosh\left(\frac{\theta +
\theta^*}{\sqrt{2}v}\right)
\right] (\mu^4 + ...)
\ee
Notice that since $K(\theta, \theta^*)$ depends only on the real component
of
$\theta$ and therefore only the real component acquires a mass! This is
completely expected, because it is not supersymmetry protecting the
imaginary
component, but the shift symmetry of the goldstone boson. Indeed, any
field
$\Phi$ whose \kp is a function of $\Phi+ \Phi^\dagger$ alone has this same
shift
symmetry in the \kp and will not acquire a mass. However, for an ordinary
field,
such a form for the \kp would seem an unnatural tuning, whereas for a
goldstone
boson it is automatic.

We will consider a theory with a spontaneous breaking of an approximate
little
$U(1)$. All terms in the superpotential will respect a spontaneously
broken
$U(1)$, but in combination they will not.

For the moment, we will not concern ourselves with the origin of this
model, and
will only address questions of technical naturalness. Later, we will
comment on
microscopic origins for the form given. The complete model is given by the
superpotential
\be
W = X_1 (\Phi \bar \Phi - v^2) + X_2 ( \frac{\kappa}{2} S^2 - \mu^2) +\lambda S^2 (\Phi +
\bar \Phi)
\ee
The equality of the $\Phi$ and $\bar \Phi$ couplings can be justified by a
$\Phi
\leftrightarrow \bar \Phi$ exchange symmetry. Notice that each term
respects a
$U(1)$ under which $\Phi$ is charged, but in combination, there is no
preserved
$U(1)$. At a scale $v$, $\phi$ and $\bar \phi$ acquire vevs and we are
left with
a non-linear sigma model below this scale. We can thus instead study the
theory
\be
W=X_2 (\frac{\kappa}{2} S^2 - \mu^2) +  2 \lambda v S^2 \cosh\left(
\frac{\Theta}{\sqrt{2}v} \right)
\ee
This model has a potential (neglecting supergravity contributions)
\be
V_0 = |\frac{\kappa}{2}s^2-\mu^2|^2 + |\kappa x_2 s + 4 \lambda v s \cosh\left(
\frac{\theta}{\sqrt{2} v}  \right)|^2 + 2 \lambda^2(s s^*)^2
\frac{|\sinh \left(\frac{\theta}{\sqrt{2}v}
  \right)|^2}{\cosh\left(\frac{\theta+\theta^*}{\sqrt{2} v}\right)}
\ee

Of course, because $\phi$ is just a pseudo-goldstone boson, radiative
effects
will give it a mass. There are two sources for such terms.  The first is due to the fact that the form of the \kp is unstable against radiative corrections, leading to induced $\phi^\dagger
\bar
\phi+h.c.$ K\"ahler potential terms. The contribution to the \kp will be
\be
\delta K = \frac{\bar \lambda^2 }{16 \pi^2} (\phi^\dagger \bar \phi+\hc),
\ee
where $\bar \lambda^2=\lambda^2 \log(\Lambda/v)$.
The $O(v^2)$ masses will drive $s$ to
$0$, while supersymmetry breaking masses for $\theta_{r}$ will drive it to
zero. 

Another contribution to the potential for $\theta_{i}$ will come from the Coleman-Weinberg potential \cite{Coleman:1973jx}. This is found to be
\bea
\delta V = \frac{\kappa^2 \mu^4}{128 \pi^2}\big[(x^2 - 1)^2 \log\left((x^2 - 1) \kappa \mu^2/\Lambda^2 \right) + \\ \nonumber
(x^2 + 1)^2 \log\left((x^2+1) \kappa \mu^2/\Lambda^2 \right) -2 x^4 \log \left( x^2 \kappa \mu^2/\Lambda^2 \right) \big]
\eea
where $x= 4 \lambda v \cos(\theta_{i}/\sqrt{2}v )/\sqrt{\kappa} \mu$. Note that one cannot estimate which contribution is more relevant merely by which is larger because the slow roll conditions involve the derivatives of $\theta_{i}$'s potential, not its absolute value. 

We will first consider each contribution in turn. If we assume the supergravity contribution is dominant, we have
\bea
V&=&\exp\left( \frac{\bar \lambda^2   v^2}{8 \pi^2 \mpl^2}  \cos \left(
\frac{\sqrt{2} \theta_{i} }{v} \right)\right)\mu^4\\
&\simeq& \mu^4+\frac{\bar \lambda^2   v^2 \mu^4}{8 \pi^2 \mpl^2} \cos
\left(
\frac{\sqrt{2} \theta_{i} }{ v} \right) .
\eea
Note that this regardless of the particular value of $\theta_{i}$, the
second term
is a negligible contribution to the total energy. We can now calculate the
slow
roll parameters
\bea
\epsilon &=& \frac{ \mpl^2}{2} \left( \frac{V'}{V} \right)^2 = 
	\frac{\bar \lambda^4 v^2}{(8 \pi^2)^2 \mpl^2} 
	\sin^2\left(\frac{\sqrt{2} \theta_{i}}{v} \right) \\
\eta &=& \frac{\mpl^2 V''}{ V} =  
	-\frac{\bar \lambda^2 }{4 \pi^2} 
	\cos \left(\frac{\sqrt{2} \theta_i}{v} \right),
\eea
where we assume an arbitrary starting value for $\theta_i \sim v$. Because
the
potential for the slow roll field is generated at the loop level, both
$\eta$
and $\epsilon$ are small. Indeed, $\epsilon$ is small even if $v \sim
\mpl$.

The additional requirements on inflation are also easily satisfied. The
total number of e-foldings during inflation is given by
\be
N =\int_{\phi_i}^{\phi_f}  \mpl^{-2} \frac{\mu^4}{V'} d \phi
\simeq \frac{4 \pi^2}{\bar \lambda^2} \log \left( \frac{\tan \theta_f}{\tan \theta_i}\right) \simeq \frac{4 \pi^2}{\bar \lambda^2} \log\left( \frac{4 \lambda v}{\sqrt{\kappa} \mu} \right),
\ee
requiring $\bar \lambda^2 \sim  O(1)$ to achieve 60 e-foldings, a very mild
requirement. Note that this requirement combined with our expressions for the slow roll parameters tells us that we would expect a blue spectrum if the SUGRA terms dominate the inflation.

COBE constraints on density perturbations require the relationship
between the scale of inflation and the symmetry breaking scale
\be
\mu^2 = 10^{-5} (\bar \lambda^2 ) v \mpl.
\ee
Since we require $v > \mu$ in order to achieve a workable hybrid inflation
model, this implies
\be
v \ge  \times 10^{-5} (\bar \lambda^2) \mpl \approx (\bar \lambda^2 )\  2
\times 10^{13} \gev,
\ee
where this limit is saturated in the limit that $v = \mu$. Note that in
the other limit, $v = \mpl$
we derive
\be
\mu \approx \bar 8\  \lambda \times 10^{15} \gev.
\ee

Now let us assume the Coleman-Weinberg contribution is dominant. Then we have slow roll parameters
\bea
\epsilon &\simeq& \frac{\mpl^2 \kappa^4 \tan^2(\sqrt{2} \theta_i/ v)}{1024 \pi^4 v^2}, \\
\eta &\simeq& -\frac{\mpl^2 \kappa^2 \sec^2(\sqrt{2} \theta_i/ v)}{14 \pi^2 v^2}.
\eea
We thus have $\epsilon/\eta = \kappa^2 \sin^2 (\theta/\sqrt{2} v) /64 \pi^2$, so $|\eta| > \epsilon$. Consequently, we would expect a red spectrum if the CW potential dominates the dynamics of inflation.

The number of efoldings is
\be
N \approx \frac{16 \pi^2 v^2}{\mpl^2 \kappa^2} \log(x_i),
\ee
hence for large $v$, order one $\kappa$ will suffice, while in general we require $\kappa \sim v/\mpl$ to achieve the necessary inflation.
COBE requires
\be
\mu^2 v \simeq 2.3 \times 10^{-6} \mpl^3 \kappa^2 \tan \left(\sqrt{2} \theta_i/v \right).
\ee

In conclusion, in both scenarios (supergravity dominated and Coleman-Weinberg dominated) the loop factor is responsible for both the slow
roll and
the large number of e-foldings. The density perturbations make a
requirement on
the scales in the theory. However, while $\mu \ll v \ll \mpl$ is
consistent with
the theory, we do not require multiple hierarchies in the theory. Both
$\mu \sim
v$ and $v \sim \mpl$ are consistent scenarios with only two scales
present and no other very small parameters, and no other small parameters at all if $v \sim \mpl$.

\subsection{Sequestered Realizations}
The theory presented is technically natural, in the sense that some
couplings are included only at their radiatively induced size. It is
interesting to consider whether there are microscopic realizations which
set these couplings to zero at tree level automatically. Here we will discuss the use of
sequestering to achieve this, while in section \ref{sec:gauge} we will
consider models where the inflaton is part of a higher dimensional
gauge field.

While the precise model described above is difficult to realize on the
basis of
symmetries and sequestering, a very similar model can be. Let us consider
a five
dimensional theory on $S_1/Z_2$ with a gauged $U(1)$ propagating in the
fifth
dimension. Let us further suppose that we spontaneously break the $U(1)$
on two
boundaries by fields $\Phi_{1,2}, \bar \Phi_{1/2}$ with charges $\pm 1$,
and
that a bulk hypermultiplet $S, \bar S$ propagates in the fifth dimension
and has
charge $\pm 1/2$ under the $U(1)$. Then we can go to the gauge where
$\Phi_1 =
\bar \Phi_2 = \Phi = e^{\theta/2 v}$ and $\bar  \Phi_1 = \Phi_2 = \bar
\Phi =
e^{-\theta/2 v}$ .

The superpotential is constrained because no direct couplings between
$\phi_1$
and $\phi_2$ can occur due to locality. However, we can write

\be
W \supset X (S \bar S-\mu^2) + \bar S^2 (\Phi_1 + \Phi_2) + S^2 (\bar
\Phi_1 + \bar \Phi_2),
\ee
which, going to the nonlinear sigma model yields
\be
W \supset X(S \bar S - \mu^2) + \bar S^2 (\Phi +\bar  \Phi) + S^2 (\bar
\Phi + \Phi).
\ee
This is a minor modification of the earlier model, and the subsequent
analysis
follows with only trivial changes.

\section{Inflation from Extra Dimensions}
\label{sec:gauge}

Now we present models of the inflaton as a component of a higher-dimensional
gauge field.
In a five-dimensional theory with a $U(1)$ gauge symmetry, $A_5$, the fifth component of the gauge boson, is protected by a remnant of the full gauge symmetry 
and therefore its mass is cut off by the compactification scale, making it a perfect candidate
for an inflaton.  It is the Wilson line which is simply a pNGB as is clearly shown by deconstruction \cite{Arkani-Hamed:2001ca}.  Finally, we present a supersymmetric version of gauge inflation 
which is to some extent a ``reconstructed'' version of the model in section \ref{sec:susy}.
We briefly describe  a six-dimensional version with $A_6$ as the waterfall field, and its deconstruction.  Adding compact extra dimensions requires their stabilization, which we shall assume happens independent of these models.  This assumption is of course not required
in the deconstructed case.  Note that components of gauge fields in extra (and deconstructed)
dimensions has been suggested as a mechanism for quintessence \cite{Hill:2002kq}.

\subsection{Gauge Inflation}

The Lagrangian of a $U(1)$ gauge theory in five dimensions compactified on a circle,
with a charged scalar $\sigma$ in the bulk is
\begin{equation}
{\cal L}_5  = |\partial_M \sigma + i g_5 A_M \sigma |^2 + {1\over 2} F^{MN} F_{MN} + V(|\sigma|^2)
\end{equation}
where $M,N=0,1,2,3,5$ and the Lagrangian has dimensions of (mass)$^5$.
The field $A_5(x,y)$ will have an initial non-zero value before the 60 e-foldings.
Five-dimensional gauge invariance allows one to transform a general profile
for the field in the fifth direction to one which is flat (independent of $y$.
This is the Wilson line which, below the compactification scale $M_c\equiv R^{-1} = (2\pi L)^{-1}$, is
simply a scalar field.  This is the scale at which loop corrections to the zero-mode
($A_5^0$) mass are cut off. This motivates us to consider $A_5\rightarrow \phi$ as a slow roll field. 

Let us begin by considering the simplest theory with one extra dimension and a charged scalar field. We will take the fundamental scale $M_*$ to be a free parameter, up to the requirement that $M_* > L^{-1}$. If the one extra dimension is the only parametrically large dimension, then we have the relation $\mpl^2 = M_*^3 L$ \cite{Arkani-Hamed:1998rs}. However, if there are other extra dimensions in which the fields we consider do not propagate, the fundamental scale can be lower.

The potential we will consider will be
\be
V=\lambda (|\sigma|^2 - v^2)^2 + g^2 |\sigma|^2 \phi^2
\ee

Radiative effects tell us that the natural size are $\lambda \gsim g^4/16 \pi^2$ and $v^2 \sim M_*^2/16 \pi^2$ if $\lambda>g^2$ and $g^2 M_*^2/\lambda 16 \pi^2$ if $\lambda < g^2$. Note that we have not yet included any potential for $\phi$. 

The potential for $\phi$ is a subtle issue. Because $\phi$ is identified with the fifth component of a gauge field in an extra dimension, the dynamics which generates it arises only from non-local interactions. Specifically, it will arise from loops of charged fields which propagate all the way around the extra dimension. However, these loops will be exponentially suppressed if the masses are large. For $\sigma$, which have a negative mass-squared, in order to have a $\sigma =0$ stable during inflation, we will need to take $|m^2_\sigma|< 1/g L$.

One can calculate the effective potential exactly, as the Lagrangian is quadratic in $\sigma$.
\bea
\delta V &=& \int \frac{d^4 q}{(2\pi)^4} \sum_{n=-\infty}^{\infty} \log[(n + g \phi R)^2 + (q R)^2 + (m R)^2]
\\ \nonumber &=& \int \frac{d^4 q}{(2\pi)^4} \log \left ( \frac{\sin(i f \pi- g \phi \pi R)\sin(i f \pi + g \phi \pi R)}{\sin^2(i \pi f)}\right),
\eea
where $f =R \sqrt{m^2 + q^2}$.

When the waterfall field has a small mass compared to the compactification scale, this is well approximated by a cosine potential of the form
\be
\delta V \approx A (1-\cos(  2 g \pi R \phi))
\ee
where 
\be A = \frac{93 \zeta(5)}{1024 R^4 \pi^6} \simeq \frac{3}{32 R^4 \pi^6}.
\ee
The slow roll parameters are given by
\bea
\epsilon &=& \frac{9 g^2 \mpl^2}{512 \pi^{10} \lambda^2 v^8 R^6} \sin^2 (2 g \pi R \phi) \\ \nonumber
\eta &=& \frac{3 g^2 \mpl^2}{8\lambda v^4 \pi^4 R^2} \cos (2 g \pi R \phi).
\eea
The total efoldings are
\be
N = \frac{8 \lambda \pi^4 R^2 v^4}{3 g^2 \mpl^2} \log \left( \frac{\tan(g \pi R \theta_i)}{\tan(g \pi R \theta_f)} \right).
\ee

There are two interesting limits we can take, one in which the vev of $\sigma$ is as small as possible and one in which it is as large as possible ($\lambda v^4 \sim 1/( R)^4$). We can reexpress $\mpl^2 = M_*^3 2 \pi R V_n$, where $V_n$ is the additional volume in fundamental units. 

In the small vev case we have
\bea
\epsilon &=& \frac{2304 \sin^2(2 g \pi R \phi) V_n}{g^2 M_*^5 R^5 \pi x^4}, \\ \nonumber
\eta &=& \frac{192\pi \cos(2 g \pi R \phi) V_n}{M_* R x^2},
\eea
where $x = g/\sqrt{\lambda} \ \ ( \sqrt{\lambda}/g)$ if $g^2 > \lambda$  ($\lambda > g^2$).
If we take a large energy density (of order the compactification scale) we have
\bea
\epsilon &=& \frac{9 g^2 (M_* R)^3 \sin^2(2 g \pi R \phi) V_n}{256 \pi^9}, \\
\eta &=& \frac{3 g^2 \pi (M_* R)^3 \cos(2 g \pi R \phi) V_n}{4 \pi^3}.
\eea

Irrespective of the other parameters of the model, we have from COBE  
\be
\eta \simeq 83 \cos(2 g \pi R \phi)\left( \frac{g^4}{\sin^2(2 g \pi R \phi)} \right)^{1/3},
\ee
so we need a somewhat small $g$ in order to satisfy density pertubations.

\subsection{Pseudo-pseudo:  the waterfall field}

It is possible to make both the slow-roll and waterfall fields components of
gauge fields and thus have a theory without fundamental scalars.
By going to six dimensions, the job of the waterfall field can be done by
$A_6$.  We will take a compactification scale for the sixth dimension higher than
that for the fifth dimension, {\it i.e.}, $L_6^{-1} > L_5^{-1}$.

Start with a six-dimensional $SU(2)$ gauge theory compactified with boundary
conditions along the sixth dimension which break the $SU(2)$ to $U(1)$.
This can be done by an orbifold such that under the transformation $x_6\rightarrow - x_6$,
the $A_{\mu}^0, A_5^0, A_6^{\pm}$ are even and the  $A_{\mu}^{\pm}, A_5^{\pm}, A_6^0$
are odd.  Only the even fields have zero mode and so below the scale $L_6^{-1}$,
there is a five-dimensional $U(1)$ gauge field and a charged scalar which we
identify as $\sigma \equiv A_6^{+}/\sqrt{L_6}$.

The necessary coupling between inflaton and waterfall appears automatically as a result
of gauge invariance.  The gauge kinetic term contains $F_{56} F^{56}$ which in turn contains terms $\tr{[A_5^0,\sigma^{\pm}]^2}$.  In addition, a potential for $\sigma$ is generated below
$L_6^{-1}$.  Thus the inflaton-waterfall coupling is built in!  All that is needed is a potential
for the waterfall field.  This can be generated by a fermion doublet living on a boundary of the 
fifth dimension (say at $x_5=0$), or living in the full bulk with a mass 
$m_f\sim L_6^{-1}\gg L_5^{-1}$.  The fermion will couple to $A_6$ through normal gauge
interactions and give a negative squared mass and positive quartic term to the waterfall
field, similar to the top contributions to the Higgs effective potential in the standard model
of particle physics.  Below the compactification scale of the sixth dimension, the dynamics
are as in the five-dimensional model of the previous section.

We can deconstruct the six-dimensional model above, {\it i.e.}, latticize the
two extra dimensions.  However, we can retain the salient features of the model
above with very few lattice sites:  two to be precise.

The model is of the same form as the ``minimal moose'' \cite{Arkani-Hamed:2002qx}
model used to explain a light Higgs and electroweak symmetry breaking.  There are 
four sets of link fields:
\begin{equation}
\Phi_i = f e^{i {\tilde\theta}_i /f} = f e^{i \theta^a \tau^a /f}
\end{equation}
which each parameterize the
breaking of $SU(2)_L\times SU(2)_R \rightarrow SU(2)_V$, just like the pions of QCD.
Two $U(1)$ symmetries are gauged, $U(1)_L \times U(1)_R$, such that all links transform
as
\begin{equation}
\Phi_i \rightarrow e^{- i \tau^3 \alpha(x)_L} \Phi_i e^{i \tau^3 \alpha(x)_R} \ .
\end{equation}
In other words, we have gauged a $U(1)$ subgroup of the four $SU(2)_L$s and similarly
for the $SU(2)_R$.  This is the equivalent to orbifolding.  Adding the operators
\begin{equation}
g_1 \tr{\Phi_1 \Phi_2^\dagger \Phi_3 \Phi_4^\dagger} + g_2 \tr{\Phi_1 \Phi_4^\dagger \Phi_3 \Phi_2^\dagger} 
\end{equation}
produces a quartic potential mixing the neutral and charged components of the $\Phi\,$s.
Both couplings are required to produce a quartic-type coupling at tree level, and both are
required to break the symmetries of some of the fields, thus suppressing one-loop contributions
to their masses.  Using the $U(1)$ gauge couplings as spurions, it is possible to introduce or generate
a suitable potential for all of the fields, allowing for technical naturalness to suppress the 
sizes of operators.

While we do not explore the full details of this model here, we note that it would be of 
interest to produce a ``moose'' model \cite{Georgi:1986hf}
from a theory of fermions and gauge fields and without
fundamental scalars.

\subsection{Supersymmetry in Extra Dimensions}
These extensions to higher dimensional gauge theories also work well in
the supersymmetric models.
Again, let us consider a
five-dimensional $U(1)$ gauge theory with a bulk hypermultiplet $S, \ \bar
S$. The action for the bulk fields \cite{Arkani-Hamed:2001tb} (including
the radion
\cite{Marti:2001iw}) are (we follow the notation of \cite{Kaplan:2001cg})
\bea
S_5 &=& \int d^4 x \ d\phi \left[ \int d^2 \theta \frac{T}{4 g_5^2}
W^\alpha
  W_\alpha +\bar S(\partial_\phi - \frac{\chi}{\sqrt{2}}) S+ \hc \right]
\\
\nonumber &+&
\int d^4\theta \frac{1}{T+T^\dagger} \left[ \frac{2}{g_5^2} \left(
\partial_\phi
  V - \frac{1}{\sqrt{2}}\left(\chi+\chi^\dagger \right) \right)^2
  \right]+\frac{T+T^\dagger}{2}(S^\dagger S + \bar S^\dagger \bar S).
\label{eq:5dmodel}
\eea
We then add a boundary term which explicitly breaks the $N=2$
supersymmetry and the $U(1)$
\be
S_4 = \int d^4 x \ d\phi \delta(\phi) \int d^2 \theta X (\frac{\kappa}{2} S \bar S -
\mu^2).
\ee
Notice that once the radion is stabilized to a value $R$, this reduces to
a a
field with the appropriate Yukawa superpotential term, and most
importantly, a
kinetic term that depends only on $\chi+\chi^\dagger$. This is a remnant
of the
shift symmetry of the theory which protects the field from getting a mass.
Of
course, below the compactification, there is no remnant of the higher
dimensional gauge theory and radiative effects should spoil this as in the
little inflaton model.

The fact that an $N=2$ gauge theory coupled to charged matter has the
proper form for a hybrid inflation model has been noted by \cite{Watari:2000jh} and
\cite{Kallosh:2001tm} and later
studied by \cite{Kallosh:2001gr}. These models were 
not little inflaton models, however, because the \kp employed was
minimal, and they did not arise from fundamentally higher dimensional gauge theories. There the supergravity $\eta$ problem was solved by assuming the absence of Planck scale vevs in \cite{Watari:2000jh} and was solved by driving
inflation with Fayet-Iliopoulos terms D-terms in \cite{Kallosh:2001tm}. F-term inflation would
still have induced a large mass in this latter case. 

In our model, the particular form  of the \kp is
enforced by the five dimensional gauge symmetry, but below the
compactification scale, radiative effects will spoil this. Moreover, contributions from the Coleman-Weinberg potential will also give a mass to $\chi$. 

The corrections to the \kp will be
\be
\delta K = -\frac{g^2}{16 \pi^2} \log(\vev{\chi} L) \chi \chi^\dagger,
\ee
while the Coleman-Weinberg contributions will be 
\bea
\delta V &=& \frac{\kappa^2 \mu^4}{64 \pi^2}\Big[ -2 x^4 \log(\frac{x^2 \kappa \mu^2}{\Lambda^2}) + (x^2-1)^2 \log(\frac{(x^2-1)\kappa \mu^2}{\Lambda^2})\\ \nonumber &&+(x^2+1)^2 \log(\frac{(x^2+1)\kappa \mu^2}{\Lambda^2})\Big]
\eea
where $x = g \chi/\sqrt{2 \kappa} \mu$. This yields for  $x \gg 1$
\be
\delta V \approx \frac{\kappa^2 \mu^4}{32 \pi^2} \log(x^2 \kappa \mu^2/\Lambda^2).
\ee
Again we consider each case in turn, first when supergravity corrections are dominant and then when Coleman-Weinberg contributions are dominant.

We begin with supergravity contributions, which give a mass term
\be
\delta V \approx -\frac{g^2 \mu^4}{16 \pi^2 \mpl^2}\log(\vev{\chi} L) \chi \chi^*.
\ee
The slow roll parameters are then
\bea
\epsilon = \frac{g^4}{128 \pi^4 \mpl^2}  \chi \log(\vev{\chi}L)^2,\\
\nonumber
\eta = -\frac{g^2}{8 \pi^2} 2 \log(\chi L).
\eea
Thus, slow roll is easily achieved, even with $O(1)$ $g$. The number of efoldings is
\be
N=\frac{8 \pi^2}{g^2} \log( \log (\chi_i L)/\log(\chi_f L) ),
\ee
while COBE requires
\be
\mu \approx 6\times 10^{16} \gev \frac{g}{2 \pi} \sqrt{\chi \log(\pi/L \chi)/2 \sqrt{2} \mpl}.
\ee

If the Coleman-Weinberg potential drives the slow-roll field, this essentially the model of \cite{Dvali:1994ms}, and we have slow roll parameters
\bea
\epsilon&=& \frac{g^2 \kappa^3 \mpl^2 x^2}{4 (4\pi)^4 \mu^2} (x^2 \log\left(\frac{x^4}{x^4-1}\right) + \log\left(\frac{x^2-1}{x^2+1}\right))^2
,\\ \nonumber
\eta &=&-\frac{g^2 \kappa \mpl^2}{32 \pi^2 \mu^2}(3 x^2 \log\left(\frac{x^4}{x^4-1}\right)+\log\left(\frac{x^2-1}{x^2+1}\right)).
\eea
For $x\gg1$ these simplify to
\bea
\epsilon&=& \frac{g^2 \kappa^3 \mpl^2 }{4 (4\pi)^4 \mu^2 x^2} ,\\ \nonumber
\eta &=&-\frac{g^2 \kappa \mpl^2}{32 \pi^2 \mu^2 x^2}.
\eea

Note that we have $\epsilon = \kappa^2 \eta/32 \pi^2$. So merely by perturbativity ($\kappa < 4\pi$) we have a $|\eta| > \epsilon$. The number of efoldings is
\be
N= \frac{16 \pi^2 \mu^2 x_i^2}{ g^2 \kappa \mpl^2},
\ee
so even for $O(1)$ $\kappa$, $g$, we can have sufficient inflation.
Given the relatively simple form for $N$, we can rewrite the slow roll parameters at the COBE era as
\bea
\epsilon \simeq \frac{\kappa^2}{64 \pi^2 N_{COBE} },\\ \nonumber
\eta \simeq \frac{-1}{2 N_{COBE}}.
\eea

 Lastly COBE requires
\be
\mu = 6 \times 10^{16}\gev \sqrt{\frac{k}{8 \pi \sqrt{N_{COBE}}}}.
\ee

In conclusion, the corrections to the potential are sufficiently soft that slow roll is easily achieved, along with satisfying COBE requirements, all without any very large or small parameters.

\section{Implications of WMAP}
\label{sec:WMAP}
With the release of the new WMAP data \cite{Bennett:2003bz,Spergel:2003cb}, cosmology has moved into a new era of precision data. In particular, we have a great deal of information now on the tilt parameter $n$, and tantalizing hints for possible non-zero $dn/d\log k$. A broad discussion of the implications for inflation is given in \cite{Peiris:2003ff}.

Let us begin by simply summarizing the WMAP results: from WMAP alone, $n=0.99 \pm 0.04$, while inclusion of 2dFGRS \cite{Colless:2001gk,Percival:2001hw} and studies of Lyman $\alpha$ systems \cite{Croft:2000hs,Gnedin:2001wg} gives $n=0.93 \pm 0.03$. From this we see the data prefer red spectra, but are still consistent with blue spectra. These results assume a running spectral index, which is found to have $dn/d\log(k) = -0.031^{+0.016}_{-0.017}$. This suggests that $n$ is moving red on small distance scales, and a particularly remarkable indiciation is that the spectrum actually moves from blue ($n>1$) on long distance scales to red ($n<1$) on short distance scales.
In light of this, we will examing the models studied here, beginning with the non-supersymmetric models and moving on to the supersymmetric ones. 

The ``littlest inflaton'' model of section \ref{sec:littlest} has very small couplings, which are a priori independent. Without any explanation of their origin, one requires a fine-tuning of parameters in order to achieve any observable deviation from $n=1$. However, if such a tuning did occur, one would require the log enhancement of equation \ref{eq:e-folds} (that is, $\tan(2 \theta_i/f) \gg1$), and hence $\eta < 1$ would be possible, allowing a red spectrum.

Next we consider the non-supersymmetric ``gauge inflation'' model. Here we have the relation 
\be
\epsilon \approx \frac{3}{64 \lambda \pi^6 R^4 v^4} \eta
\ee
Hence, if the inflation scale ($\lambda v^4$) is small compared to the compactification scale, it is natural to have $\epsilon > |\eta|$, yielding a red spectrum. However, if $|\eta| > \epsilon$, because the potential involves a cosine, $\eta$ can have either sign. As a consequence it is difficult to make predictions with this model. However, $dn/d\log k$ can only be negative if $\epsilon$ is significant (yielding possibly observable tensor perturbations and a red spectrum) or in the top half of the potential (where the curvature is negative). Hence, if the observations of a spectrum changing from blue to red are verified, this non-supersymmetric model will be excluded.

The supersymmetric models are perhaps the most interesting because they have {\em two} sources to their potential: one from the Coleman-Weinberg potential and one from supergravity corrections. This is analagous to the situation in \cite{Linde:1997sj}, with two important differences: first, we have no need to make additional assumptions about the short-distance physics of the theory. Unlike \cite{Linde:1997sj}, we do not need to assume small vevs for all moduli fields, and all dangerous higher dimension operators which could give a large mass to the slow roll field are absent at tree level. This arises from the shift symmetry of the goldstone boson or the five-dimensional gauge symmetry. Secondly, the theory is entirely predictive from the renormalizable parameters of the theory - the non-renormalizable operators which are significant are those that arise out of the tree-level supergravity potential.

This being said, let us study the supersymmetric models in more detail. We will focus on the extra dimensional model, while the deconstructed model is quite similar with $1/L$ identified with $v$. Recall that in the CW regime, $\eta \approx -1/2 N$. The transition between the CW and the SUGRA regime (when $V_{SUGRA}'=V_{CW}'$ and $V_{SUGRA}''=V_{CW}''$) occurs at $\chi= \mpl \kappa/\sqrt{2} g$ (here we ignore the logarithmic dependence of the SUGRA potential, keeping only the dominant power behavior and setting $\bar g \sim g$). The efoldings in the CW regime is then 
\be
N_{CW} = \frac{4 \pi^2}{g^2}.
\ee
Hence, for order one $g$, the transition from SUGRA to CW naturally occurs at $O(40)$ efoldings before the end of inflation. As pointed out in \cite{Linde:1997sj}, this is interesting because the transition from SUGRA to CW regimes gives a transition from a blue to a red spectrum, which is a feature suggested at $2 \sigma$ from the WMAP data. However, unlike \cite{Linde:1997sj} this is quite natural with order one parameters, and does not require additional small parameters.

In summary, the analyzed models are consistent with the WMAP data, although the ``littlest inflaton'' would appear finely tuned if there is increasing certainty of $n \ne 1$. The non-supersymmetric gauge inflation model can satisfy a red spectrum with non-zero $dn/d\log k$, but cannot accomodate a change from blue to red. If the data continue to support this, this model will be excluded. The supersymmetric models quite naturally give all the qualitative features observed. A change from blue to red is quite natural in the regime suggested by WMAP, and, in fact arises in the model without any fine tuning of the parameters, or any assumptions about the coefficients of non-renormalizable operators. A complete study of these models is therefore warranted.

\section{Discussion}
\label{sec:conc}
Inflation solves the big problem of the flatness of the universe, leaving a little problem
in its wake -- the flatness of the inflaton potential.  This new class of particles -- pNGBs
with collective symmetry breaking -- fits perfectly in this role.  

It is clear now that fundamental scalars are not required for inflation.
Higher components of extra-dimensional gauge fields can contain
all of the required dynamics.  Little inflaton models can be generated by
confining theories of fermions and gauge fields,  possibly even
to construct renormalizable ones ({\it i.e.}, no n-fermion operators required).

For pNGBs in supersymmetric theories, the $\eta$-problem is solved.
The kinetic term for these fields always come in the form $T + T^\dagger$,
and any modulus could potentially play the role of the inflaton.  For example,
if $T$ were the radion multiplet in a five-dimensional supersymmetric theory,
its imaginary scalar component is the fifth component of the gravi-photon $B_5$.
Thus it may be that a model which stabilizes the compact dimension already has all of 
the ingredients to produce inflation.

Previous discussions of pNGBs have involved either axion-like potentials (so-called ``natural inflation'') or situations where the approximate symmetry was very good, in that the explicit symmetry breaking parameters were very small. Here, we need not rely on extremely high scales (as in natural inflation) or very small parameters. Moreover, because the dynamics is more complicated, there is a greater hope for observable consequences in the CMBR.

\vskip 0.2 in
\noindent{\bf Acknowledgments}
The authors would like to thank Andy Cohen for very useful conversations on inflation and Nemanja Kaloper for reading the draft and useful comments, as well as Ann Nelson and Raman Sundrum for useful discussions. The authors would also like to thank the theory group at Boston University where much of this work was undertaken.
\vskip 0.15in
\noindent{\it Note added:} As this work was being completed, we became aware of 
\cite{Arkani-Hamed:2003wu}, which also considers the $A_5$ as a candidate for the inflaton, as well as work in progress by the same authors on similar ideas.

\bibliography{li}
\bibliographystyle{apsrev}

\end{document}